\documentclass{osa-article}

\journal{oe}

\begin{document}

\title{Overlapping two standing-waves in a microcavity for a multi-atom photon interface}

\author{S\'{e}bastien Garcia,\authormark{1,2,$\dagger$} Francesco Ferri,\authormark{1,3} Jakob Reichel,\authormark{1} and Romain Long \authormark{1,*}}

\address{\authormark{1} Laboratoire Kastler Brossel, ENS-Universit\'{e} PSL, CNRS, Sorbonne Universit\'{e}, Coll\`{e}ge de France, 24 rue Lhomond, 75005 Paris, France}
\address{\authormark{2} Present address: Young Physics team's, Institute of Physics, Coll\`{e}ge de France, 11 place Marcelin Berthelot, 75005 Paris, France}
\address{\authormark{3} Present address: Institute for Quantum Electronics, ETH Z\"{u}rich, 8093 Z\"{u}rich, Switzerland}

\email{\authormark{$\dagger$}sebastien.garcia@college-de-france.fr,\authormark{*}long@lkb.ens.fr} 

\begin{abstract}
We develop a light-matter interface enabling strong and uniform coupling between a chain of cold atoms and photons of an optical cavity.  This interface is a fiber Fabry-Perot cavity, doubly resonant for both the wavelength of the atomic transition and for a geometrically commensurate red-detuned intracavity trapping lattice. Fulfilling the condition of a strong and uniform atom-photon coupling requires optimization of the spatial overlap between the two standing waves in the cavity. In a strong-coupling cavity, where the mode waists and Rayleigh range are small, we derive the expression of the optimal trapping wavelength taking into account the Gouy phase. The main parameter controlling the overlap of the standing waves is the relative phase shift at the reflection on the cavity mirrors between the two wavelengths, for which we derive the optimal value. We have built a microcavity optimized according to these results, employing custom-made mirrors with engineered reflection phase for both wavelengths. We present a method to measure with high precision the relative phase shift at reflection, which allows us to determine the spatial overlap of the two modes in this cavity.    
\end{abstract}

\section{Introduction}

The development of light-matter interfaces has played a key role in the progress achieved in laser physics, nonlinear optics and quantum optics. 
In recent years, the emergence of quantum technologies underlines the need of developing new interfaces for applications ranging from quantum communication to quantum metrology. In Cavity Quantum Electrodynamics (CQED)~\cite{Haroche2006}, the light-matter coupling is enhanced by placing emitters inside a cavity. One prominent milestone in this field has been the achievement of the strong coupling regime between a single emitter and a cavity mode \cite{Meschede1985, Rempe1987, Brune1996, Thompson1992, Reithmaier2004a, Yoshie2004, Wallraff2004}.

Focusing on atomic systems, single atoms strongly coupled to a cavity are envisioned as the elementary nodes of future quantum networks~\cite{Ritter2012, Gallego2018}. The interaction between a cold atomic ensemble and a cavity mode in the weak coupling regime at the single atom level has led to the generation of multiparticle entangled states useful for metrology~\cite{Schleier-Smith2010, McConnell2015, Hosten2016}. 
Light matter interfaces have also been pushed to explore the regime where each atom of an atomic ensemble is strongly coupled to the cavity mode.  This allows to engineer a strong controllable effective coupling between the atoms mediated by the cavity field, enabling the generation of specific multiparticle entangled states~\cite{Barontini2015, Welte2017} and the simulation of many-body models~\cite{Baumann2010, Landini2018, Kroeze2018a, Davis2019}.

For experiments with many atoms, the question arises on how to implement a strong and yet uniform coupling between the cavity mode and each atom of the atomic ensemble, which requires that all the atoms interact with the same maximal value of the cavity field. 
To implement this condition, one approach is to minimize the size of the atomic sample by using Bose-Einstein Condensates (BEC) and loading the BEC inside a single antinode of the cavity mode~\cite{Colombe2007, Brennecke2007}. This method is very effective for generating collective interaction but it is not compatible with single particle detection and control, which has emerged as a powerful tool in the fields of trapped ions~\cite{Blatt2008}, Rydberg atoms~\cite{Deleseleuc2019} and ultracold atoms in optical lattices~\cite{Bakr2009a}. In the context of CQED, single atom addressability has only been realized in systems with two atoms~\cite{Reimann2015,Welte2017}.

\begin{figure}[thbp]
\centering
\includegraphics[width=0.95\textwidth]{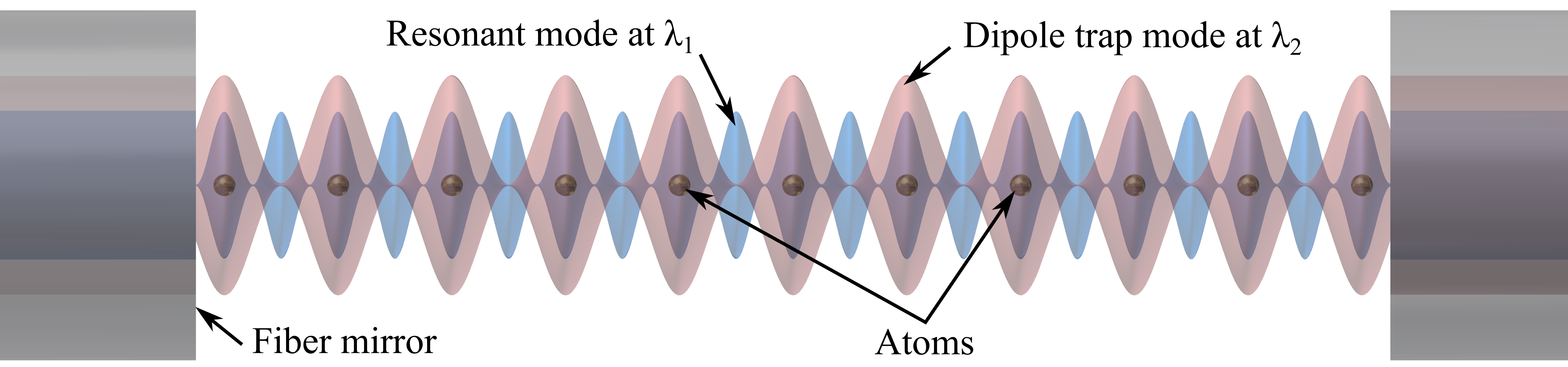}
\caption{Schematic view of our multi-atom photon interface in a fiber Fabry-Perot cavity (not to scale). The photonic mode (blue) of the quantum interface has a wavelength $\lambda_1$ close to the atomic resonance. A second auxiliary cavity mode (red) with wavelength $\lambda_2 \simeq 2 \lambda_1$, red-detuned from atomic resonance, creates an array of microtraps that overlap with each second maximum of the field at $\lambda_1$. This optimizes the coupling between the atomic array and the cavity photons at $\lambda_1$.}
\label{fig:AtomCavityScheme}
\end{figure}

In order to combine single particle resolution and a strong uniform coupling of each atom with the cavity mode resonant with the atomic transition, we use the spatial periodicity of the standing waves of linear Fabry-Perot cavity modes. By precisely positioning single atoms at different antinodes of the cavity mode, we can obtain a one-dimensional array of single atoms with maximal uniform coupling and single particle control. To get a stable overlap between the atoms and the cavity mode at the wavelength $\lambda_{1}$ of the atomic transition, we aim to produce such a chain of atoms by optically trapping the atoms at the antinodes of a second red-detuned cavity mode at a commensurate wavelength $\lambda_{2} \simeq 2 \lambda_{1}$, as presented in Fig.~\ref{fig:AtomCavityScheme}. Such a configuration has been used previously to couple atomic ensemble to a macroscopic cavity in the weak coupling regime~\cite{Vrijsen2011, Arnold2012, Lee2014a, Kollar2015, Davis2019}. 

In this article, we show that extending this scheme to atomic ensembles where each atom is strongly coupled to the cavity field necessitates a careful maximization of the spatial overlap between the two standing-waves of the doubly-resonant cavity. We demonstrate how to optimize this overlap, we present experimental methods to measure it and we apply them to our doubly-resonant cavity. The overlap is especially critical for CQED experiments in the strong coupling regime, as they usually require cavities with high finesse and small mode volume, which leads to the natural choice of microcavities such as fiber Fabry-Perot cavities~\cite{Hunger2010}. For such short cavities, the Rayleigh range is usually small and so the Gouy phase shift plays a major role. In addition, the atomic ensemble spreads over a large fraction of the cavity length (on the order of the Rayleigh length), which imposes special constraints for the optimization. 
A similar problematic arises in frequency doubling by using a doubly-resonant cavity, where a good spatial overlap is also required between the fundamental mode and the second harmonic one \cite{Ashkin1966, Wu1985, Paschotta1994a, Liscidini2006a}. But, microcavities are usually not used in this context, which relaxes the need for an advanced optimization of the overlap.

In section \ref{text:dual-wavelength theory}, we analyze theoretically the parameters that affect the overlap and we derive conditions to optimize it. In particular, we show in the following that the naive condition $\lambda_{2} = 2\lambda_{1}$ needs corrections due the geometric Gouy phase shift which leads to a non-trivial optimal wavelength for the trapping light. Given this condition, we demonstrate that the maximal overlap is obtained for an optimal relative phase shift between the two wavelengths at the reflection on the cavity mirrors. We illustrate the effect of this optimization by plotting the characteristic overlap phase and the strength of the coupling between a single atom and a photon for all trapping positions along the cavity axis. In Sec.~\ref{tex:coating}, we discuss the implementation of the optimal relative phase on reflection for our dielectric mirror coatings. Then, we report on experimental techniques which we developed to measure this relative phase. First in Sec.~\ref{text:Tip}, we observe directly the overlap via the perturbation of the two modes at $\lambda_{1}$ and $\lambda_{2}$ with a nanoscopic probe and we deduce a coarse value of the relative phase. Second, in Sec.~\ref{text:rel_phase}, we present a measurement that allows us to determine the relative phase with high precision, by measuring the double resonance condition as a function of the cavity length.

\section{Doubly-resonant cavities with maximal spatial overlap of the standing waves}
\label{text:dual-wavelength theory}

In this first part, we analyze and optimize the parameters that determine the spatial overlap of the modes in order to maximize it. We consider a linear Fabry-Perot resonator whose geometrical length is given by $L$, formed by two identical spherical concave mirrors, i.e. with identical radius of curvature $R$ and identical coating. This configuration places the waist of the cavity modes at the center of the resonator and thus maximizes here the coupling to emitters. The following analysis can be extended to the case of asymmetric cavities where the two mirrors have different radii of curvature. 

\subsection{Optimal wavelength}

In the simple case of plane waves, the maximal spatial overlap between the standing waves at $\lambda_{1}$ and $\lambda_{2}$ is simply obtained by the condition $\lambda_{2} = 2\lambda_{1}$. For Gaussian cavity modes, we only consider in the following the fundamental transverse Gaussian mode TEM$_{00}$, so transverse coordinates are not involved. However, the Gouy phase shift has to be taken into account, which is the additional phase shift accumulated by  a Gaussian beam when passing through its waist compared to an ideal plane wave. As a consequence, the effective wavelength of the Gaussian beam around the focus can be significantly different from $c/ \nu$, where $\nu$ is the laser frequency and $c$ the speed of light. 
In order to obtain a maximal spatial overlap between the two wavelengths $\lambda_{1}$ and $\lambda_{2}\approx 2\lambda_{1}$, this local variation of the periodicity has to be taken into account to determine the optimal wavelength.  By defining $z$ as the coordinate along the cavity axis and by setting $z=0$ as the position of the cavity waist, the phase of a mode propagating along the cavity axis is given by $\frac{2\pi}{\lambda}z - \Phi_{\mathrm{G}} (z)$, where $\Phi_{\mathrm{G}}(z)=\arctan \left( \frac{z}{z_{\mathrm{R}}}\right)$ is the Gouy term with $z_{\mathrm{R}}=\frac{1}{2}\sqrt{L(2R-L)}$ the Rayleigh length. At the first order, the effective wavelength at the position $z$ is then given by:
\begin{equation}
\lambda^{\mathrm{eff}}_{i}(z)=\left(\frac{1}{\lambda_{i}}-\frac{1}{2\pi}\frac{d\Phi_{\mathrm{G}}}{dz}(z)\right)^{-1} \ \ .
\label{eq:conditionoverlap}
\end{equation}
A maximal overlap is then obtained by the condition $\lambda^{\mathrm{eff}}_{2}(z)=2\lambda^{\mathrm{eff}}_{1}(z)$.

This condition cannot be fulfilled for all $z$ because the Gouy term is independent on the wavelength. By choosing to impose this condition at the waist ($z=0$) for a symmetric cavity, we obtain a relation between the wavelengths of the two modes: 
\begin{equation}
\lambda_{2}^{\mathrm{opt}}=\frac{2\lambda_{1}}{1+\frac{\lambda_{1}}{2\pi z_{\mathrm{R}}}}  \ \ .
\label{eq:wavelength}
\end{equation}
This simple relation guarantees a quasi-optimal overlap between the two effective wavelengths over a length smaller or comparable to $z_{\mathrm{R}}$ around the waist.

For our specific experiment, $\lambda_{1}$ is the wavelength of the laser probe, which is fixed by the Rubidium atomic transition to $780.24\,$nm (D$_2$ line of $^{87}$Rb).  The cavity mirrors have a radius of curvature $R\simeq 300\,\mu$m  and the cavity length is $L\simeq130\,\mu$m. From Eq.~\ref{eq:wavelength}, we then get $\lambda_{2}^{\mathrm{opt}} \simeq 1558.92\,$nm, which is significantly different from  $2 \lambda_{1}$. This is due to the small value of the Rayleigh length $z_R$ and the difference $\lambda_{2}^{\mathrm{opt}}-2 \lambda_{1}$ can exceed the tunability range of typical diode lasers.  Using this value for $\lambda_{2}$, we also calculate in the extreme off-centered position $z=65\,\mu$m, the ratio $\lambda_{2}^{\mathrm{eff}}/\lambda_{1}^{\mathrm{eff}}\simeq 1.99956$, which indicates a good match between the effective wavelengths over the whole cavity length.

\subsection{Doubly-resonant cavities}
\label{text:2lambdacav}

The cavity has to be simultaneously resonant at the two different wavelengths $\lambda_{1}$ and $\lambda_{2}$. The eigenfrequencies of such a resonator are fixed by the condition that the phase shift accumulated in one cavity round-trip has to be an integer multiple of $2\pi$:
\begin{equation}
\label{eq:1mode_resonance}
2\pi \nu_{qlp} \frac{2L}{c}-2\phi_{\mathrm{m}}-4(l+p+1)\Phi_{\mathrm{G}}\left(\frac{L}{2}\right)=2\pi q
\end{equation}
where $\nu_{qlp}$ is the frequency of the eigenmode with longitudinal order $q$ and whose Hermite-Gauss transverse shape is identified by $l$ and $p$. As we will focus on the fundamental transverse mode TEM$_{00}$, we set  $l= p=0$. $\phi_{\mathrm{m}}$ is the phase shift due to the reflection on a mirror, whose value is determined by the structure of the reflective coating. $\Phi_{\mathrm{G}}\left(\frac{L}{2}\right)=\arctan\left(\frac{L}{2 z_{\mathrm{R}}}\right)$ is the Gouy phase-shift accumulated from the center of the cavity to a mirror. One should note that the terms $\phi_{\mathrm{m}}$ and $\Phi_{\mathrm{G}}\left(\frac{L}{2}\right)$, which are fixed for a given cavity, can significantly affect the resonant frequencies in the case of microcavities (where $q$ is small).

For a doubly-resonant cavity, the condition of Eq.~\ref{eq:1mode_resonance} has to be fulfilled by both wavelengths:
\begin{eqnarray}
\label{eq:2mode_resonance}
\nonumber 2 L\frac{2 \pi}{\lambda_{1}}-2\phi_{\mathrm{m1}}-4\Phi_{\mathrm{G}}\left(\frac{L}{2}\right)&=2\pi q_{1}\\
2L\frac{2\pi}{\lambda_{2}}-2\phi_{\mathrm{m2}}-4\Phi_{\mathrm{G}}\left(\frac{L}{2}\right)&=2\pi q_{2}
\end{eqnarray}
where we have considered only the fundamental TEM$_{00}$ modes with longitudinal index $q_1$ for $\lambda_1$  and $q_2$ for $\lambda_2$. $\phi_{\mathrm{m1}}$ and $\phi_{\mathrm{m2}}$ represents the phase shifts after a reflection on the mirror for the wavelengths $\lambda_1$ and $\lambda_2$, respectively.

\subsection{Optimal relative phase}
\label{text:opt_phase}

The condition $\lambda^{\mathrm{eff}}_{2}(z)=2\lambda^{\mathrm{eff}}_{1}(z)$ is necessary but not sufficient to obtain a maximal spatial overlap between the standing waves of the two cavity modes. Indeed, even if the effective wavelengths $\lambda^{\mathrm{eff}}_{1}$ and $\lambda^{\mathrm{eff}}_{2}$ are commensurate, the relative position of the nodes of the standing waves inside the cavity is not univocal. 

\begin{figure}[thbp]
\centering
\includegraphics[width=0.8\textwidth]{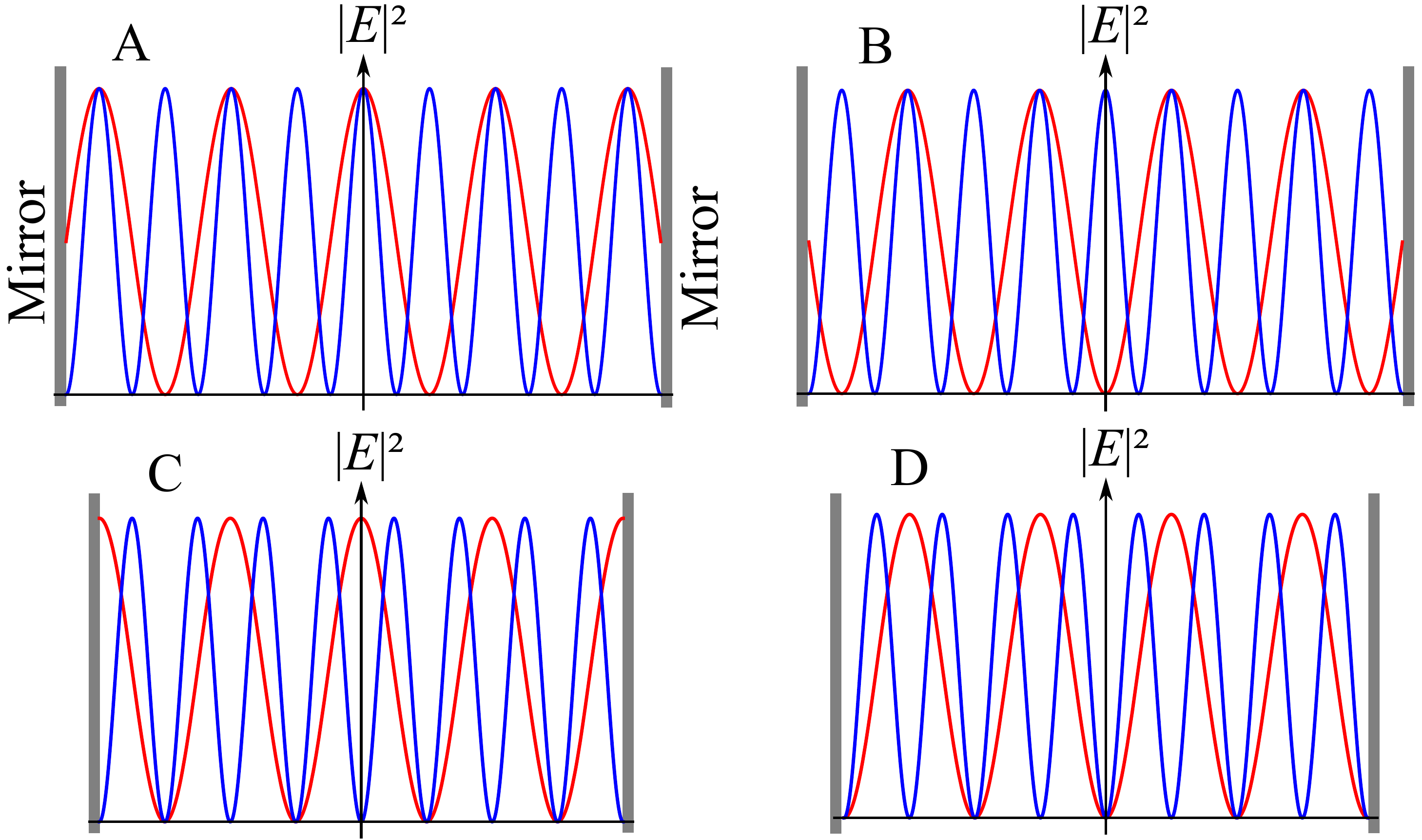}
\caption{Schematic view of the cavity field intensities $|E|^2$ for the wavelengths $\lambda_{1}$ (blue) and $\lambda_{2}$ (red).  Assuming that the two wavelengths differ by a factor $2$, we show the four possible spatial configurations (A to D) that are allowed by symmetry in a cavity with planar mirrors.}
\label{Standwave_phase}
\end{figure}

We first consider the simple configuration of a cavity with planar identically-coated mirrors which forces the condition $\lambda_{2}=2\lambda_{1}$. The symmetry imposes that each standing wave must have either a node or an antinode in the center of the cavity. Thus, only four configurations are possible, as shown in Fig.~\ref{Standwave_phase}. The targeted configurations A and B, where the antinodes overlap, are only obtained if $\lambda_{1}$ has an antinode in the center of the cavity. 
Whether the nodes of the standing waves (minimum overlap) or their antinodes (maximum overlap) are coincident is determined by the relative phase shift at reflection between the two wavelengths. In the following, we define the relative phase shift as  $\Delta\phi_{\mathrm{m}} = \phi_{\mathrm{m2}}-\frac{\phi_{\mathrm{m1}}}{2}$. As shown by the difference between the field intensities on the mirror for $\lambda_{2}$ and $\lambda_{1}$, the four configurations have indeed different relative phase shifts at reflection.
 
By using the double resonance condition of Eqs.~\ref{eq:2mode_resonance}, we get the expression of the relative phase shift at reflection:
\begin{equation}
\label{rel_phase}
\Delta \phi_{\mathrm{m}} \equiv \phi_{\mathrm{m2}}-\frac{\phi_{\mathrm{m1}}}{2} =\pi \left(\frac{q_1}{2}-q_2\right)+2\pi L \left(\frac{1}{\lambda_{2}} - \frac{1}{2 \lambda_{1}}\right) - \Phi_{\mathrm{G}}\left(\frac{L}{2}\right) \ \ .
\end{equation}
The four solutions presented in Fig.~\ref{Standwave_phase} are defined by the four integer and half-integer values of the quantity $\frac{q_1}{2}-q_2$ modulo $2$. Indeed, for a cavity with planar mirrors, the last two terms in Eq.~\ref{rel_phase} cancel because we impose $\lambda_{2} = 2 \lambda_{1}$ from Eq.~\ref{eq:wavelength} and $\Phi_{\mathrm{G}}\left(\frac{L}{2}\right)=0$. The configurations presented in Fig.~\ref{Standwave_phase} assume $\phi_{\mathrm{m1}}=\pi$ to obtain a node on the mirror, and so $q_{1}+1$ is the number of antinodes of the standing wave at $\lambda_{1}$ in the cavity. Thus, there is an antinode in the center of the cavity when $q_1$ is even, which corresponds to the configurations A and B. As a consequence, the maximal overlap of antinodes is obtained for $\Delta \phi_{\mathrm{m}} = \left(\frac{q_1}{2}-q_2 \right) \pi$  with $\frac{q_1}{2}-q_2$ integer. The solutions with $\frac{q_1}{2}-q_2$ half-integer lead to overlapping nodes configurations C and D and must be rejected.

We can rewrite Eq.~\ref{rel_phase} by keeping only the optimal solutions and by reporting the condition for the wavelength $\lambda_{2} = \lambda_{2}^{\mathrm{opt}}$ of Eq.~\ref{eq:wavelength} to get the optimal relative phase shift in the general case of a cavity with identical concave mirrors:
\begin{equation} \label{opt_phase}
\Delta\phi_{\mathrm{m}}^{\mathrm{opt}}=  n\pi + \delta\phi_{\mathrm{m}}^{\mathrm{opt}} \,\, , \,\,\,\, \textrm{with}\,\,\,\,\delta\phi_{\mathrm{m}}^{\mathrm{opt}}   \equiv \left(\frac{L}{2 z_{\mathrm{R}}}-\arctan\left(\frac{L}{2 z_{\mathrm{R}}}\right)\right),
\end{equation}
where $n$ is an integer and where the non-integer part $\delta\phi_{\mathrm{m}}^{\mathrm{opt}}$ of the relative phase depends only on two parameters : the cavity length $L$ and the radius of curvature $R$ of the mirrors. This non-integer part originates from the Gouy phase after compensation of its linear effect by the choice of $\lambda_{2}^{\mathrm{opt}}$. For our specific cavity parameters, we get $\delta\phi_{\mathrm{m}}^{\mathrm{opt}}=+2.4^{\circ}$, which is a small value because the Rayleigh length $z_{\mathrm{R}} \simeq 124\,\mu$m is on the order of the cavity length $L$. 

The electric field amplitude $E_i$ of the standing-wave at $\lambda_i$ ($i={1,2}$) on the optical axis $z$ is proportional to $\cos\left(\Psi_i\left(z\right)\right)$, where the propagation phase $\Psi_i$ is given by:
\begin{equation}
\Psi_i\left(z\right) = \frac{2 \pi}{\lambda_i}\left(z-\frac{L}{2}\right)-\left(\Phi_{\mathrm{G}}\left(z\right) - \Phi_{\mathrm{G}}\left(\frac{L}{2}\right)\right) + \frac{\phi_{\mathrm{m}i}}{2} \ \ .
\label{eq:Psi_i}
\end{equation}
Thus, the overlap between the two resonant cavity modes can be quantified by the overlap phase $\Delta\Psi\left(z\right)= 2\Psi_2\left(z\right) -  \Psi_1\left(z\right)$,  which indicates the dephasing between the two standing wave patterns. 
Indeed, in case of perfect overlap, at the position of an antinode of the $\lambda_2$ standing-wave, there is an antinode of the $\lambda_1$ standing-wave, which implies that $\Psi_2$ and $\Psi_1$ are zero modulo $\pi$, as well as $\Delta\Psi$. 
On the contrary, if a node of the $\lambda_1$ standing-wave is located at an antinode of the $\lambda_2$ standing-wave, the  overlap phase is $\Delta\Psi = \pm \pi/2$ modulo $\pi$. As we seek to maintain the overlap over all the antinodes of the $\lambda_2$ standing-wave, $\Delta\Psi$ has to remain close to zero modulo $\pi$ along the cavity axis $z$. 

For a given relative phase on reflection $\Delta\phi_{\mathrm{m}} = n\pi + \delta\phi_{\mathrm{m}}$ (with $n$ integer and $\delta\phi_{\mathrm{m}} \in [-\pi/2 , \pi/2]$) and a given wavelength $\lambda_1$, the condition of double resonance of Eqs.~\ref{eq:2mode_resonance} imposes the wavelength $\lambda_2$, chosen to satisfy the condition $\frac{q_1}{2}-q_2$ integer by using $\frac{q_1}{2}-q_2 = n$ in order to be the closest of the optimal value $\lambda_{2}^{\mathrm{opt}}$ given by Eq.~\ref{eq:wavelength}. By substituting the expressions of $\lambda_1$ and  $\lambda_2$ in  Eq.~\ref{eq:Psi_i}, we deduce the following expression of the overlap phase between the two resonant cavity modes :
\begin{equation}
\Delta\Psi\left(z\right) = \frac{z}{L/2} \delta\phi_{\mathrm{m}} + n \pi + \frac{z}{L/2} \Phi_{\mathrm{G}}\left(L/2\right) - \Phi_{\mathrm{G}}\left(z\right)  \ \ .
\label{eq:OverlapPhase}
\end{equation}
We see in this equation that the overlap phase depends explicitly on $\delta\phi_{\mathrm{m}}$ and so on the relative phase. In this equation, the $n \pi$ term modulo $2 \pi$ differentiates between the solutions with antinodes overlapping at $z=0$ (type A in Fig.~\ref{Standwave_phase}) or at $z=\lambda^{\mathrm{eff}}_1/2$ (type B in Fig.~\ref{Standwave_phase}) which can both lead to optimal overlap if $\delta\phi_{\mathrm{m}} = \delta\phi_{\mathrm{m}}^{\mathrm{opt}}$. 
As the overlap only depends on $\Delta\Psi$ modulo $\pi$, we consider the case $n=0$, where $\Delta \phi_{\mathrm{m}}=\delta \phi_{\mathrm{m}}$, for simplicity in the following analysis.

\begin{figure}[htbp]
\centering
\includegraphics[width=0.7\textwidth]{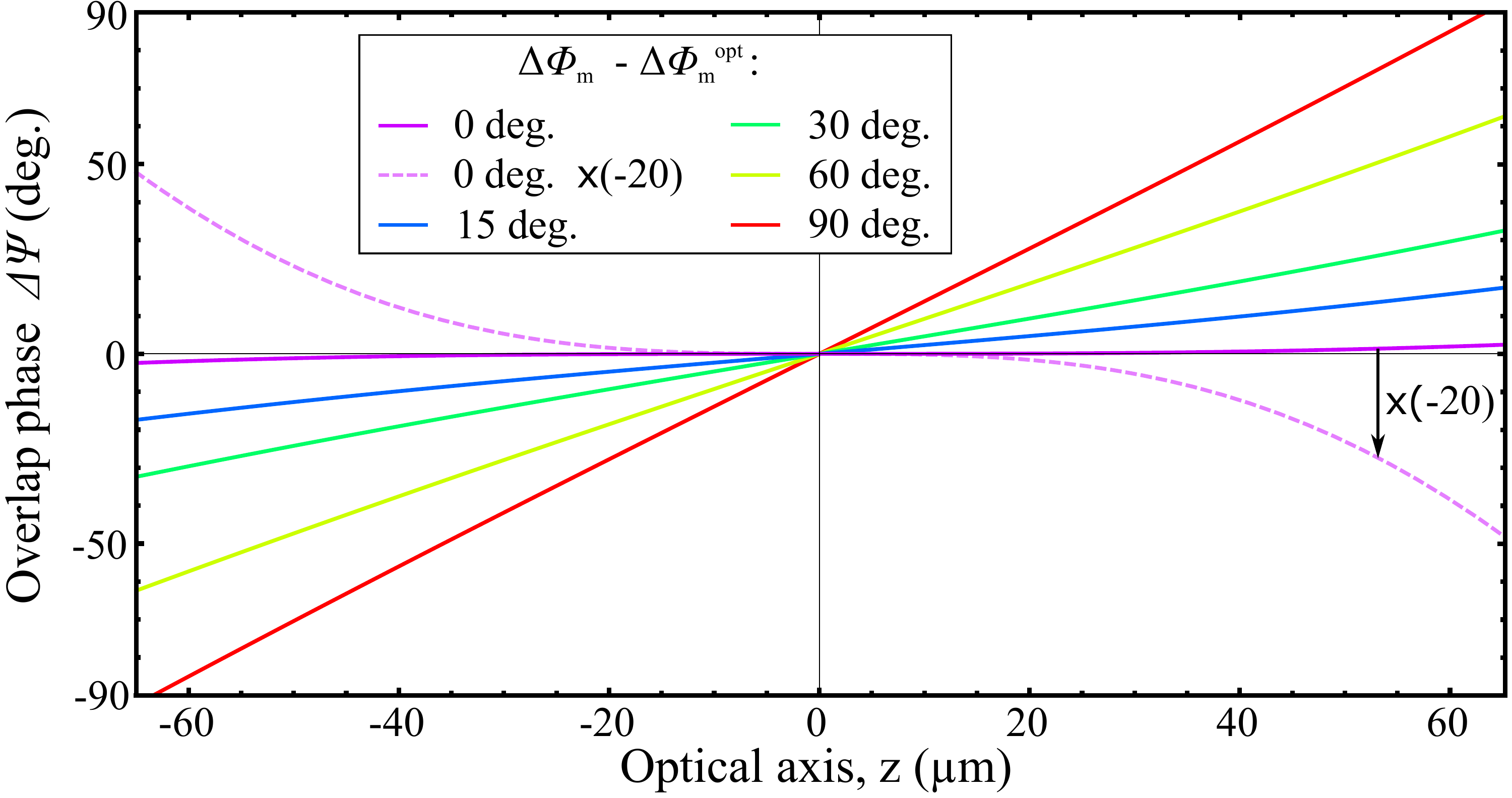}
\caption{Overlap phase $\Delta\Psi$ between the standing waves at $\lambda_{1} = 780.24\,$nm and $\lambda_{2}$ along the cavity optical axis $z$. Different colors correspond to different relative phases on reflection $\Delta\phi_{\mathrm{m}}$, compared to the optimum value $\Delta\phi_{\mathrm{m}}^{\mathrm{opt}}$. The dashed light purple curve represent the optimal-phase purple curve multiplied by a factor $-20$ to enhance the visibility.}
\label{fig:RelativePhase}
\end{figure}

In Fig.~\ref{fig:RelativePhase}, we plot the overlap phase $\Delta\Psi\left(z\right)$ for our specific cavity parameters, depending on the deviation $\Delta \phi_{\mathrm{m}}-\Delta\phi_{\mathrm{m}}^{\mathrm{opt}}$ from the optimal relative phase at reflection. In the general case where the relative phase $\Delta \phi_{\mathrm{m}}$ is not optimized, the overlap phase is dominated by the first term of Eq.~\ref{eq:OverlapPhase} and we can observe in Fig.~\ref{fig:RelativePhase} a linear variaton of $\Delta\Psi\left(z\right)$ with a slope proportional to the relative phase on reflection.
At the optimal relative phase $\Delta\phi_{\mathrm{m}}^{\mathrm{opt}}$, the overlap phase $\Delta\Psi\left(z\right)$ 
is simply given by
\begin{equation}
\Delta\Psi\left(z\right) = \frac{z}{z_\mathrm{R}} -\arctan\left(\frac{z}{z_{\mathrm{R}}}\right) + n \pi.
\label{eq:OverlapPhaseOptimum}
\end{equation}
Around $z=0$, the choice of the optimum relative phase cancels at first order the $z$ dependency of the  overlap phase, which remains close to zero over the largest possible cavity length, meaning that the antinodes of the two standing waves coincide. Even at this optimal relative phase, a perfect overlap cannot be maintained over the whole cavity range due to the change in effective wavelengths implied by the Gouy phase. However, in this case, the deviation has a zero derivative in the center of the cavity.

\subsection{Single atom-photon coupling along the atomic array}

\begin{figure}[htbp]
\centering
\includegraphics[width=0.7\textwidth]{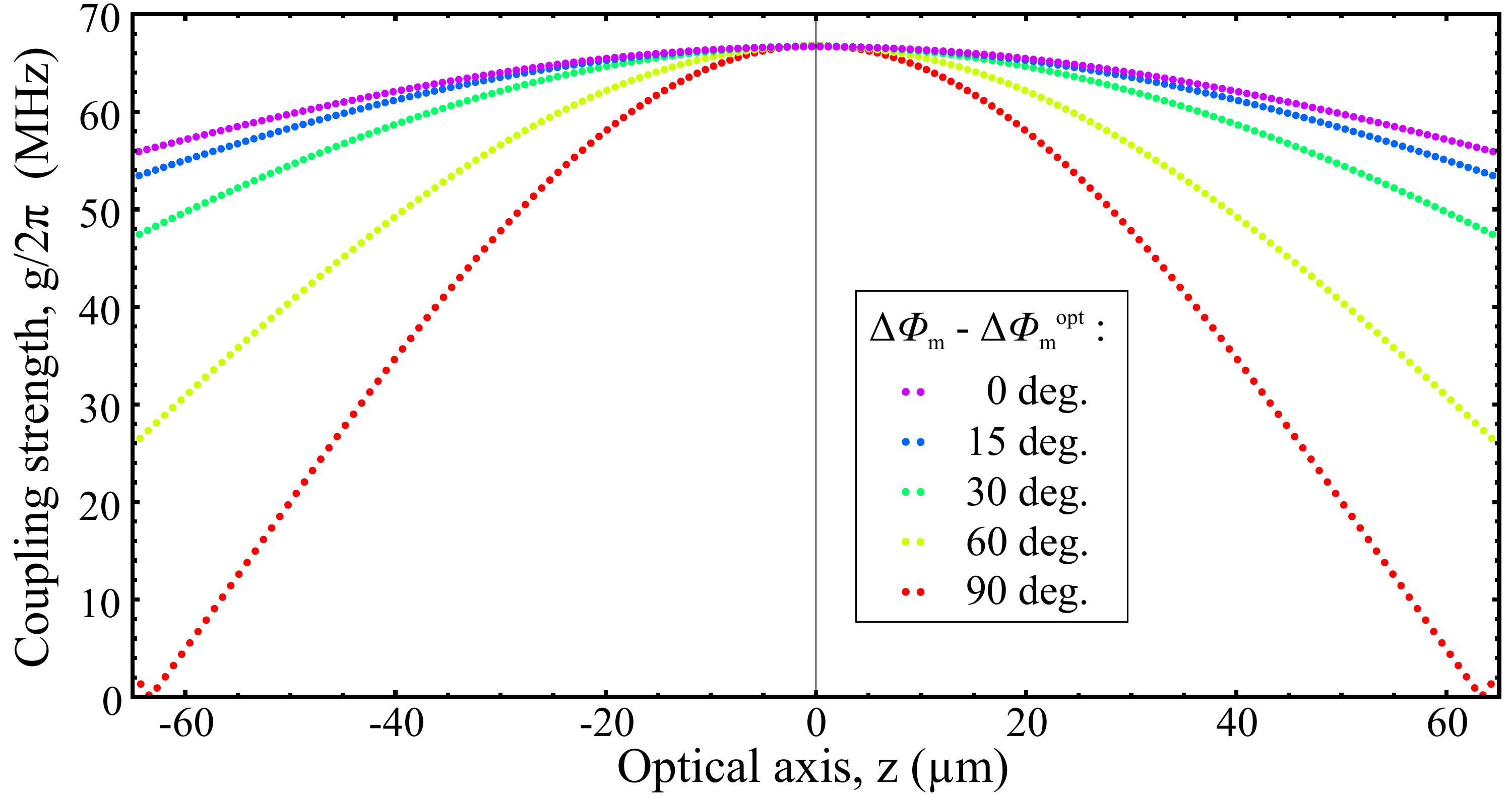}
\caption{Coupling strength $g/2\pi$ between a single atom and a single photon in the cavity field as a function of the position $z$ of the lattice site where the atom is trapped. Different colors correspond to different relative phases $\Delta\phi_{\mathrm{m}}$, compared to the optimum value $\Delta\phi_{\mathrm{m}}^{\mathrm{opt}}$. The coupling has been calculated with: a cavity length $L=130\,\mu$m, mirrors with radius of curvature $R=300\,\mu$m, a cavity finesse $\mathcal{F}_{780} = 5\cdot 10^4$, an intracavity circulating power of $100\,$mW for the dipole trap creating a maximal depth of $72\,\mu$K, and a temperature of the atom of $10\,\mu$K.}
\label{fig:Coupling_phase}
\end{figure}
 
In CQED experiments, the strong coupling regime is reached when the single atom - single photon  coupling strength $g$ is larger than the cavity decay rate $\kappa$ and the atomic one $\gamma$. For our specific atom and cavity, $\gamma/(2 \pi)\simeq 3\,$MHz and $\kappa/(2 \pi) \simeq 15\,$MHz. For an atom precisely positioned at an antinode (at the center of the cavity) of the standing wave of the resonant wavelength $\lambda_1$, we calculate the maximal value of the coupling strength $g_{max}/(2 \pi)\simeq 82\,$MHz, indicating that our system operates deeply in the strong coupling regime of CQED. 

To underline the need of carefully optimizing the overlap between the standing waves, we compute the single-atom single-photon coupling strength $g$ at each trapping site (determined by the antinodes of the standing-wave of the $\lambda_2$ mode) along the cavity axis $z$ (see Fig.\ref{fig:Coupling_phase}). 
The calculation uses our specific cavity parameters and values of the dipole trap depth and the atom temperature, that are typically used in experiments.  
We plot the coupling for different values of $\Delta\phi_{\mathrm{m}}-\Delta\phi_{\mathrm{m}}^{\mathrm{opt}}$. 
At the center of the cavity where the two standing waves are overlapped, the value of coupling strength $g/(2 \pi)\simeq 67\,$MHz is reduced compared to the maximal one due to thermal motion of the atoms in the trap wells, which leads to an averaged value of $g$.  

As the relative phase $\Delta\phi_{\mathrm{m}}$ approaches the optimal one, the variation of the coupling along $z$ becomes dominated by the divergence of the cavity mode at $780\,$nm, with a coupling difference between the center and the periphery of the cavity below $15\%$. In the opposite case, if the relative phase $\Delta\phi_{\mathrm{m}}$ deviates from the optimal one, the coupling decreases strongly while moving away from the cavity center. In the limit of a $90^\circ$ deviation, the coupling decreases to half of the maximal value at $40\,\mu$m from the center. 
This underlines the importance of implementing and verifying experimentally the conditions described above for achieving an optimal overlap between the cavity standing waves.

\section{Dual-wavelength reflective coating} 
\label{tex:coating}

As explained in the previous section, the relative phase at the reflection on a cavity mirror is a crucial parameter to obtain the maximal overlap. In this section, we discuss the implementation of the relative phase for the high reflective dielectric coating used for our fiber mirrors~\cite{Garcia2018}.

\begin{figure}[htbp]
\centering
\includegraphics[width=0.8\textwidth]{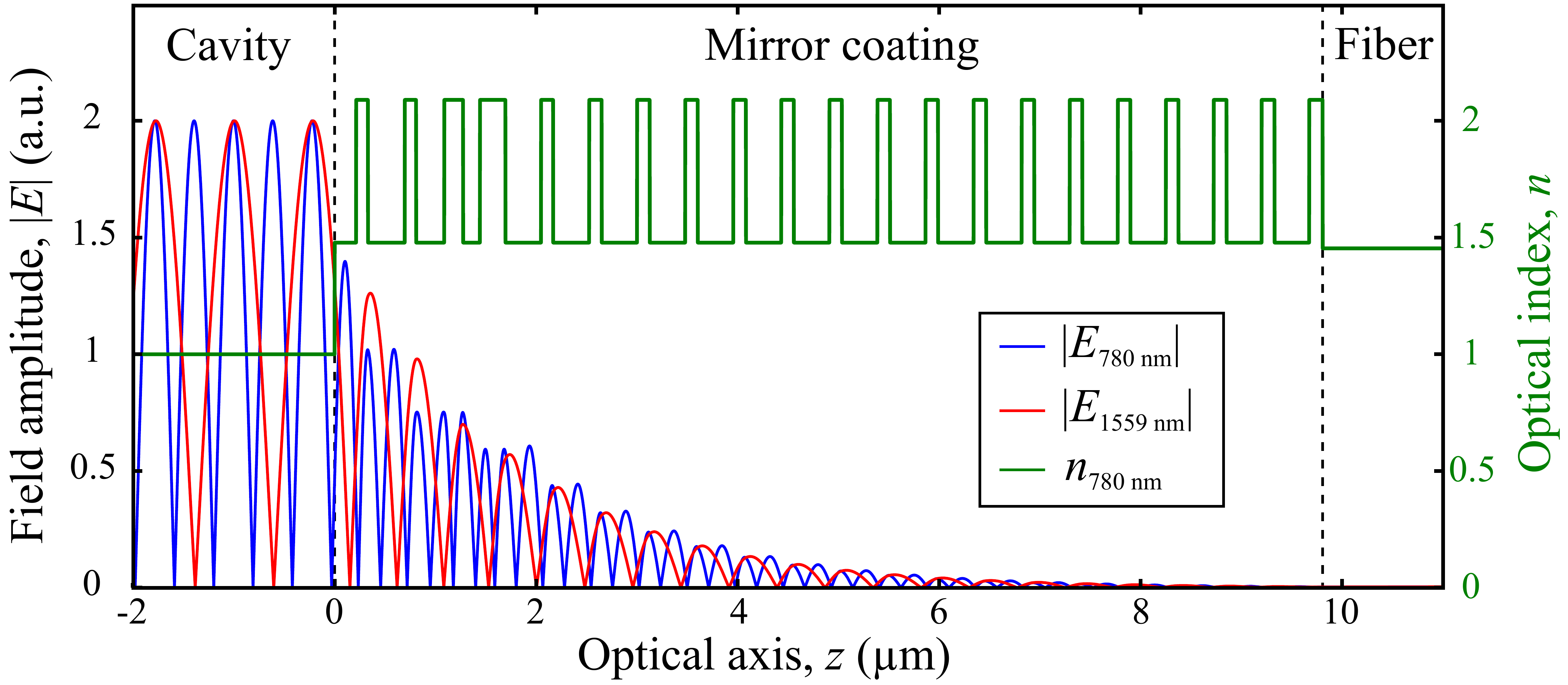}
\caption{Calculation of the mirror properties. The green curve shows the evolution of optical index $n$ along the optical axis $z$, inside the dielectric mirror (starting at $z=0$, dashed lines shows the mirror stack limits). The electric field amplitudes $|E|$ of reflected waves at $780\,$nm and $1559\,$nm are represented in blue and red, respectively. They are calculated with transfer matrix theory from the optical index and assuming an incident wave of amplitude $1$.}
\label{fig:coating_reflexion}
\end{figure}

Our mirrors are Bragg reflectors obtained by stacking layers of different dielectric materials, deposited on the end facet of the fiber by ion sputtering. The Bragg structure determines not only the global reflection and transmission coefficients of the mirrors, but also the relative phase shift at reflection between $\lambda_{1}$ and $\lambda_{2}$, which is essential for the optimal overlap between the standing waves. 
The company Laseroptik designed and realized the mirror coating taking into account the given constraint of the relative phase at reflection. The expected optical properties of our mirrors solely depend on the thickness and the optical index of the different Bragg layers, resulting in the optical index profile shown in Fig.~\ref{fig:coating_reflexion}. We performed a calculation of the optical field based on the evolution of an incident scalar plane wave in the transfer matrix formalism~\cite{Furman1992} for each wavelength $\lambda_{1}=780\,$nm and $\lambda_{2}=1559\,$nm, see Fig.~\ref{fig:coating_reflexion}. 

We observe that the fields decrease quickly while penetrating in the mirror because the layers reflect progressively the incident wave.
From the field energy distribution, we calculate the optical penetration depths $d_{\mathrm{p},1} \simeq 1.20\,\mu$m and $d_{\mathrm{p},2} \simeq 1.51\,\mu$m, at $\lambda_{1}=780\,$nm and $\lambda_{2}=1559\,$nm, respectively. The penetration of the field in the mirrors increases the effective optical length of the cavity $L_{\mathrm{eff}} = L + 2 d_{\mathrm{p}}$, which defines the free spectral range (FSR) $c/2 L_{\mathrm{eff}}$. In the calculations of the previous sections (Eq.~\ref{eq:1mode_resonance} and resulting equations), the propagation phase accumulated in the penetration depth by the wave is included in the mirror reflection phase $\phi_{\mathrm{m}}$. By using the resonance condition of Eq.~\ref{eq:1mode_resonance} to calculate the FSR and by linearizing in the FSR expression the change of the reflection phase $\phi_{\mathrm{m}} (\lambda)$ with the wavelength $\lambda$, the penetration depth can also be calculated as $ d_{\mathrm{p},i} \simeq \frac{\lambda_i^2}{4 \pi} \frac{\partial \phi_{\mathrm{m}} }{\partial \lambda} \left(\lambda_i\right)$. This calculation of the penetration depth is analytically equivalent to the one using the energy field distribution for high reflective Bragg mirrors~\cite{Babic1992}. Indeed, when applied to our calculated mirror phases at $\lambda_{1}=780\,$nm and $\lambda_{2}=1559\,$nm, this equation yields the same values of penetration depths than the ones obtained from the field energy distribution. 

In Fig.~\ref{fig:coating_reflexion}, we also directly observe that the field antinodes of the two standing waves have a very good spatial overlap. From the field calculations, we find that the fields at $\lambda_{1}$ and $\lambda_{2}$ have reflection phases $\phi_{\mathrm{m},1} = 207.6^{\circ}$ and $\phi_{\mathrm{m},2} = 99.4^{\circ}$ respectively, and thus a relative phase $\Delta\phi_{\mathrm{m}}= 0 \times \pi + \delta\phi_{\mathrm{m}}$ with $\delta\phi_{\mathrm{m}}=-4.4^{\circ}$. This value differs from the optimal value of $+2.4^{\circ}$ calculated above from our cavity parameters. The reason for this discrepancy is incidental and it originates from a different convention used by the coating company in considering the sign of the phase. However, this deviation from the optimum is small and thus the overlap is still nearly optimal, with the coupling strength at each trapping site varying by less than $0.3\%$ from the values of the optimal relative phase configuration (cf. Fig.~\ref{fig:Coupling_phase}). 

In the following paragraphs, we present two methods that allow us to confirm the calculation by measuring the relative phase shift at reflection between the cavity modes at $\lambda_{1}$ and $\lambda_{2}$. The first one is a direct visualization of the standing waves, allowing to distinguish the configurations of maximal and minimal overlap (Fig.~\ref{Standwave_phase}), \textit{i.e.} to get the integer contribution to $\Delta\phi_{\mathrm{m}}$ (modulo $2 \pi$). The second technique consists in a measurement of the precise value of the phase difference by estimating $\delta\phi_{\mathrm{m}}$.

\section{Measurement of the overlap of cavity standing waves with a tapered fiber tip}
\label{text:Tip}
 
\begin{figure}[htbp]
\centering
\includegraphics[width=0.95\textwidth]{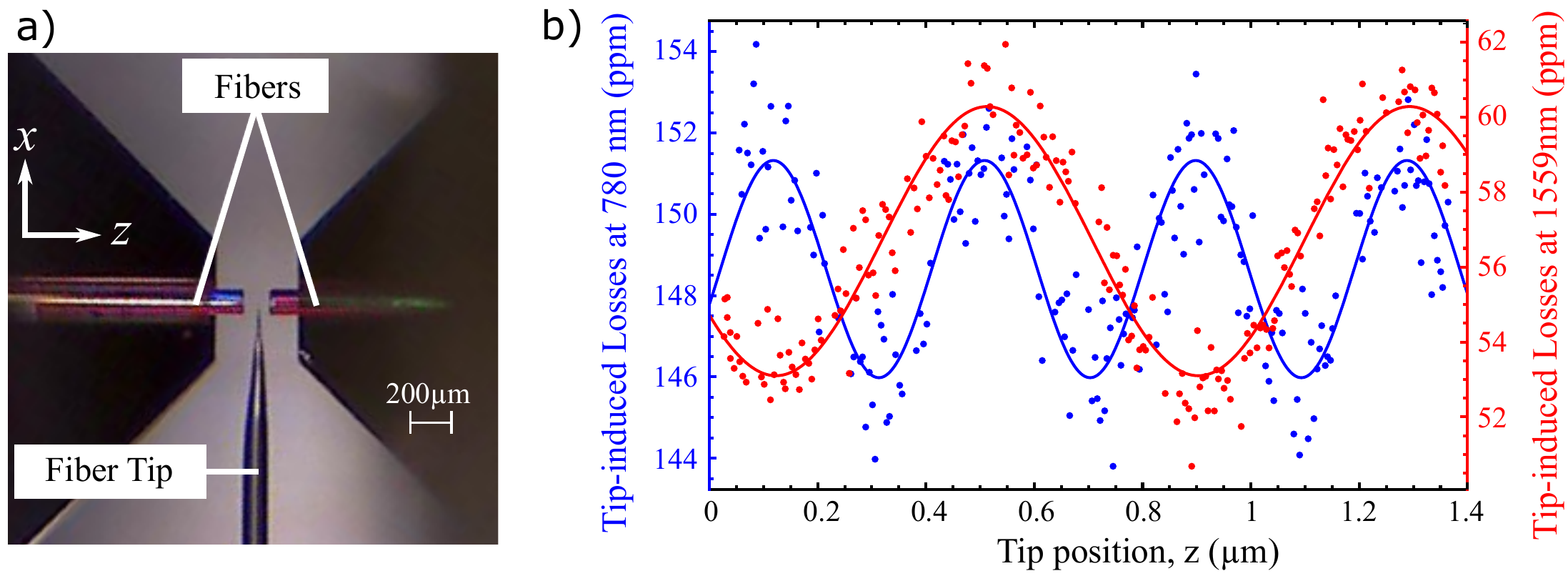}
\caption{ (a): Image of the fiber tip inserted in the fiber cavity for the measurement of the overlap between optical standing waves. (b): Tip-induced cavity losses (data points) at $\lambda_{1}$ and $\lambda_{2}$ as a function of the tip position $z$ along the cavity axis. The solid curves are best-fit sinusoids.}
\label{fig:StWave}
\end{figure}

We directly visualize the cavity standing waves with the method described in Ref.~\cite{Ferri2019}. We use a tapered fiber tip with sub-wavelength apex size to perturb the cavity modes, see Fig.~\ref{fig:StWave}(a). We measure the additional losses induced by the tip from the change of cavity transmission. 
When the tip is displaced along the cavity optical axis $z$, the tip-induced losses change periodically between a maximum value, when the tip is at the antinodes of the cavity standing waves, and a minimum value, when the tip is at the nodes.
In Fig.~\ref{fig:StWave}(b), we show the result of two sweeps of the tip along a few periods of the standing waves along the cavity axis $z$: one for each wavelength $\lambda_{1}$ and $\lambda_{2}$. The two sweeps are performed at different depth of the tip inside the mode, corresponding to the two optimal positions for the signal-to-noise, as described in Ref.~\cite{Ferri2019}.

In Fig.~\ref{fig:StWave}(b), the maxima of the losses at $\lambda_{1}$ and $\lambda_{2}$ coincide, which is the signature that the spatial configuration of the standing waves is the one maximizing the overlap between the antinodes ($\frac{q_1}{2}-q_2$ integer), and not the one where the nodes are coincident ($\frac{q_1}{2}-q_2$ half-integer). The sinusoidal fit functions give the difference between the fitted positions of the antinodes at $\lambda_{2}$ and $\lambda_{1}$ of about $1\,$nm, which is below the uncertainty of our alignment procedure of about $\pm 30\,$nm~\cite{Ferri2019}.  
Knowing that the tip was positioned at about $40\,\mu$m from the center of the cavity and that increasing $z$ corresponds to a movement towards the mirror, we obtain a rough estimation of the relative phase shift at the reflection:
\begin{equation}\label{eq:Tip_phase_results}
\Delta \phi_{\mathrm{m}}= n \pi + \delta \phi_{\mathrm{m}}\quad \mathrm{with}\quad n \,\, \mathrm{integer} \quad \mathrm{and} \quad \delta \phi_{\mathrm{m}}=(-1\pm 25)^{\circ}.
\end{equation} 
A complementary measurement, presented in the next section, allows us to estimate $\delta \phi_{\mathrm{m}}$ with a much better precision, using the result obtained here that $n$ is an integer.

\section{Precise measurement of the relative phase shift at reflection}\label{text:rel_phase}

We implement a method for measuring with a better precision the relative phase shift $\delta\phi_{\mathrm{m}}$. 
The procedure consists in measuring the laser frequency $\nu_{2}=c/\lambda_2$ for which the cavity is simultaneously resonant at $\nu_{2}$ and at  $\nu_{1}=c/\lambda_1$ (which is kept fixed) for different cavity lengths. As we will show in the following, the variation of the laser frequency $\nu_{2}(L)$ as a function of the cavity $L$ depends strongly on the relative phase at reflection.    

Starting from the condition of simultaneous resonance of Eq.~\ref{eq:2mode_resonance}, we divide by $2$ the equation for $\lambda_{1}$ and subtract it from
 the one for $\lambda_{2}$, we obtain:
\begin{equation}\label{phase_meas1}
\frac{2L}{c}2\pi\nu_{2}=\frac{2L}{c}2\pi\frac{\nu_{1}}{2}+2\Phi_{\mathrm{G}}\left(\frac{L}{2}\right)+2\Delta\phi_{\mathrm{m}}-2\pi\left(\frac{q_{1}}{2}-q_{2}\right).
\end{equation}
By inserting in this equation, the relative phase as $\Delta\phi_{\mathrm{m}} = n \pi + \delta\phi_{\mathrm{m}}$ with $n$ integer or half-integer and $\delta\phi_{\mathrm{m}} \in [-\pi/4 , \pi/4]$) , we obtain:
\begin{equation}
\label{phase_meas2}
\nu_{2}\left(L\right)= \frac{\nu_{1}}{2} + \frac{c}{2L} \frac{1}{\pi} \left( \Phi_{\mathrm{G}}\left(\frac{L}{2}\right) + \delta\phi_{\mathrm{m}} \right)
+ \frac{c}{2L} \left( n - \left( \frac{q_{1}}{2}-q_{2} \right) \right)
\end{equation}
In this expression, the last term, which contains the $n \pi$-term of the relative phase, cancels because experimentally we choose $\frac{q_{1}}{2}-q_{2}=n$ such that the frequency $\nu_2$ is the closest possible to $\nu_1/2$ to get $\lambda_2 \simeq \lambda_{2}^{\mathrm{opt}} \simeq 2\lambda_1$. 
Due to this effect, the measurement method, we will present below, only allows to determine the value of non integer part $\delta\phi_{\mathrm{m}}$ of the relative phase. As we know from the fiber tip measurement presented before that $n$ is an integer, we can determine unambiguously $\Delta\phi_{\mathrm{m}}$ modulo $\pi$, and so the overlap between the coupling and trapping standing waves. Without a direct measurement of the standing waves overlap, this method can still be used in combination with a calculation of the optical field taking account the parameters of the Bragg layers of the mirrors (see Sect. \ref{tex:coating}).

\begin{figure}[htbp]
\centering
\includegraphics[width=0.95\textwidth]{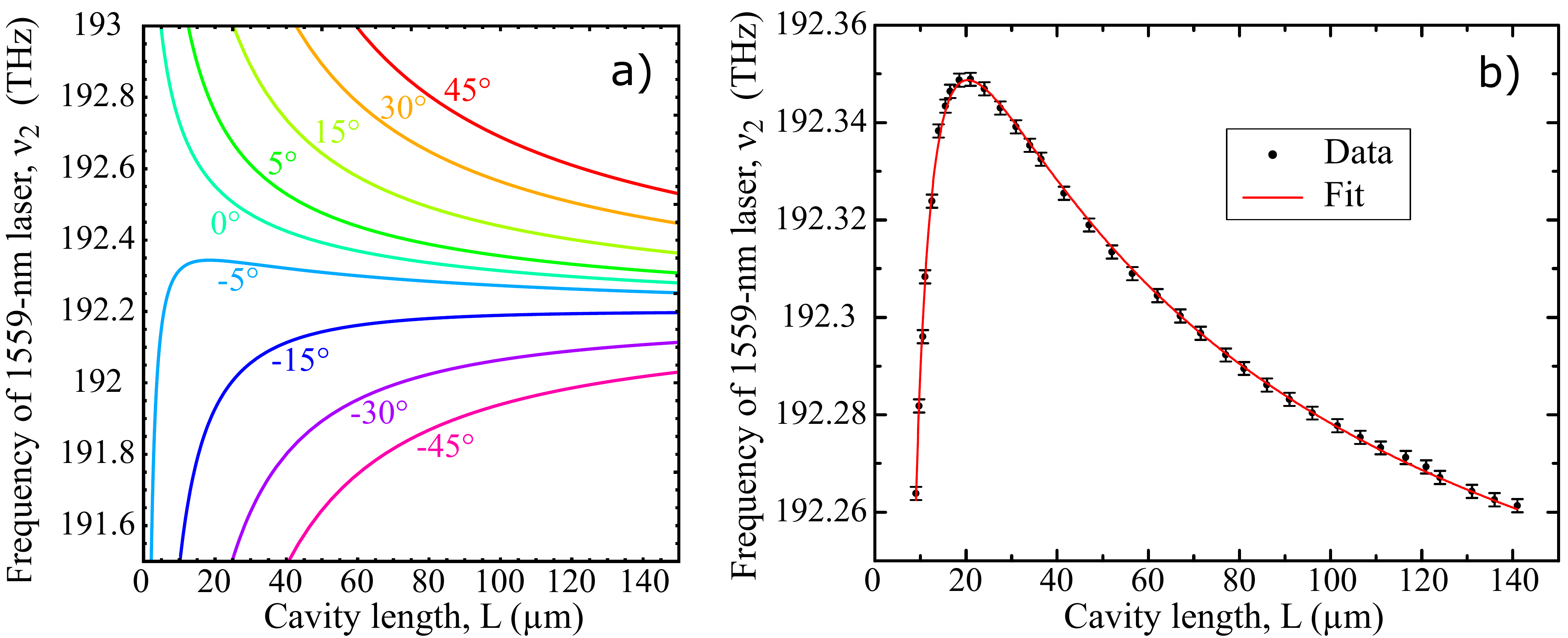}
\caption{Resonant frequency $\nu_{2}$ of the $1559\,$nm laser  for which the cavity is simultaneously resonant at $\nu_{1}$ (fixed to rubidium D$_2$ line), as a function of the cavity length $L$. (a): Theoretical curves given by Eq.~\ref{phase_meas3} for different values of $\delta\phi_{\mathrm{m}}$ (indicated in degree next to the curve). (b): Experimental data (black points) and best-fit curve (red) obtained with Eq.~\ref{phase_meas3}. The error bars on the frequency ($\pm1.4\,$GHz) represent the precision of our laser frequency calibration; the precision of the relative cavity length of $\pm0.5\,\mu$m is not shown.}
\label{Phase_measurement}
\end{figure}

By writing explicitly the expression of the Gouy phase and the Rayleigh length, we find:
\begin{equation}\label{phase_meas3}
\nu_{2}(L)=\frac{\nu_{1}}{2}+\frac{c}{2L}\frac{1}{\pi}\left(\mathrm{arctan}\left(\sqrt{\frac{L}{2R-L}}\right)+\delta\phi_{\mathrm{m}}\right)
\end{equation}
This variation of the frequency $\nu_2$ necessary to maintain the double resonance condition while keeping $\nu_1$ fixed is plotted in Fig.~\ref{Phase_measurement}(a) as function of the cavity length $L$ for different values of $\delta\phi_{\mathrm{m}}$. We clearly observe that small differences in the value of $\delta\phi_{\mathrm{m}}$ lead to strongly different curves, thus allowing for high precision measurement of the relative phase shift at reflection. 

Experimentally, the frequency $\nu_{1} = 384.228\,$THz is fixed by tuning a $780\,$nm laser diode to a given transition of a saturated-absorption spectroscopy signal of the rubidium D$_2$ line. The frequency $\nu_{2}$ of a $1559\,$nm laser diode is controlled by adjusting the current or the temperature of the laser diode.  
To observe the cavity resonances, we measure the cavity transmission while modulating the cavity length with an amplitude of approximately $\lambda_1$ with a piezoelectric actuator. At the double resonance, both transmissions of the $780\,$nm laser and $1559\,$nm laser are maximum for the same voltage on the piezoelectric actuator. In order to reach this situation, we adjust the frequency $\nu_{2}$ of the $1559\,$nm laser to the closest double resonance from $\nu_1/2$. Starting with a cavity length of approximately $150\,\mu$m, the length $L$ of the cavity is progressively reduced by a micrometric screw with a resolution of $\pm0.5\,\mu$m. At every step, $\nu_{2}$ is tuned in order to be simultaneously resonant with $\nu_{1}$. In order to know the frequency $\nu_2$ of the $1559\,$nm laser, we interpolate the values of $\nu_2$ provided by a spectrometer for different values of the current and the temperature,  with a relative precision estimated to be $1.4\,$GHz. 
We repeat this procedure up to reach the shortest accessible value of the cavity length. 

We obtain the data presented in Fig.~\ref{Phase_measurement}(b), that we fit with the model of Eq.~\ref{phase_meas3}, knowing the average measured radius of curvature of the mirrors $R \simeq 300\,\mu$m.  Systematic errors on the measured values of the laser frequency $\nu_2$ and of the cavity length $L$ are taken into account by adding offsets that are free parameters for the fitting function $\nu_{2}(L+\delta L) + \delta\nu_2$ where $\nu_{2}$ is given by Eq.~\ref{phase_meas3}. The resulting fitted values $\delta\nu_2 \simeq 3.6\,$GHz and $\delta L \simeq 2.2\,\mu$m are well below the accuracies of the spectrometer of $60\,$GHz and of the cavity length estimation of $5\,\mu$m, respectively. As these additional fit parameters only lead to linear displacements of the curve, they do not affect the determination of the relative phase $\delta\phi_{\mathrm{m}}$ which controls the shape of the curve. 

For our cavity, the phase $\delta\phi_{\mathrm{m}}$ is slightly negative; thus, in the expression of $\nu_2$ (see Eq.~\ref{phase_meas3}), the term in $\delta\phi_{\mathrm{m}} / L$ is negative whereas the Gouy phase term is positive.
The term in $\delta\phi_{\mathrm{m}} / L$  determines the divergence of $\nu_2$ when L goes to zero, because it dominates the Gouy phase term which asymptotically diverges as $1/\sqrt{L}$ when $L \ll R$. Thus, with our small negative relative phase, in Fig.~\ref{Phase_measurement}(b) when the cavity length tends to zero, the frequency $\nu_2$ is first dominated by a $1/\sqrt{L}$ asymptotic behavior that leads to an increase towards higher values originating from the Gouy phase term, but eventually $\nu_2$ diverges to smaller values of $\nu_2$ due to the $\delta\phi_{\mathrm{m}} / L$ term. 

From the fit, we find that the relative phase shit at reflection is $\delta\phi_{\mathrm{m}}=(-4.9\pm0.1)^{\circ}$. This value is in good agreement with the calculation based on the coating properties, from which it deviates by only half a degree. We emphasize the excellent uncertainty of $0.1^{\circ}$ obtained by our measurement method, which is a consequence of the high precision of the laser frequency measurement and of the translation stage that adjusts the cavity length.

\section{Conclusion}

In this article, we optimize the spatial overlap between commensurate standing waves in a doubly-resonant cavity. We derive two conditions that have to be fulfilled to optimize this overlap. First, the effective wavelength (including the Gouy phase) of the trapping lattice has to be commensurate with the effective wavelength of the mode that is strongly coupled to the atoms. This implies that the wavelength $\lambda_2$ of the trapping mode has to fulfill Eq.~\ref{eq:wavelength}. We note that such an optimization is possible because we have some freedom in the choice of the the trapping wavelength $\lambda_2$, as opposed to frequency doubling experiments.  Second, the antinodes of the two standing waves have to coincide, which is critically linked to the relative phase shift $\Delta\phi_{\mathrm{m}}$ at the reflection on the cavity mirrors. We calculate the expression of Eq.~\ref{opt_phase} of the optimal value of $\Delta\phi_{\mathrm{m}}$ as a function of the cavity parameters. 

The relative phase is controlled by the design of the dielectric mirror and can be calculated with standard transfer matrix calculations. In order to experimentally check the overlap and measure $\Delta\phi_{\mathrm{m}}$, we develop a two step method. First, the visualization of the optical field distribution with the fiber tip gives a direct measurement of the spatial overlap, which allows us to verify that the maxima of the cavity fields (and not the minima) are matched. Second, this result is completed by a more precise characterization of the relative phase, and therefore of the spatial overlap, by measuring the frequency $\nu_2(L)$ of the dipole laser which satisfies the double resonance condition as a function of the cavity length $L$.

Using this procedure, we could demonstrate that a nearly-optimal spatial overlap is achieved in our cavity. This is a critical requirement to produce new atom-photon interfaces, where a strong and uniform coupling between the resonant cavity mode and each atom of the array is combined with single particle control. The spacing between trapping sites is indeed large enough to be resolvable by a high numerical aperture lens as described in Ref.~\cite{Ferri2019}. This opens the way to new CQED experiments, where collective operations mediated by the cavity and local operations on each site of the lattice can be used to generate and study multi-particle entanglement. Such experiments can address the generation and characterization of entangled states delocalized  over the entire atomic array, useful for multiparameter quantum-enhanced sensing~\cite{Baumgratz2016, Gessner2018}. The control of a single atom of the array can be used to generates Schr\"{o}dinger cat-state ~\cite{Gerry1997} or to perform controlled-string operation~\cite{Jiang2008}. Finally, the control over the local coupling at each site between the atoms via the cavity field enable the simulation of specific spin models~\cite{Davis2019, Bentsen2019a}.

\section*{Funding}
This work was supported by :\\
Agence Nationale de la Recherche (ANR) (SAROCEMA project, ANR-14-CE32-0002); European Research Council (ERC) under the European Union's Horizon 2020 research and innovation programme Grant agreement No 671133 (EQUEMI project); and the DIM SIRTEQ from R\'{e}gion Ile-de-France.\\

\section*{Acknowledgments}

The authors acknowledge Pierre-Antoine Bourdel for careful reading of the manuscript and Tobias Gross (Laseroptik) for fruitful interaction and expert advice.\\

\end{document}